\documentclass{rspublic}

\usepackage{graphicx}
\usepackage{amsmath}
\usepackage{ulem,color}

\begin{document}

\newcommand{\sam}[1]{\textcolor{blue}{#1}} 
\newcommand{\srm}[1]{\textcolor{blue}{\sout{#1}}} 
\newcommand{\david}[1]{\textcolor{red}{#1}} 
\newcommand{\drm}[1]{\textcolor{red}{\sout{#1}}} 

\newcommand{\beq}{\begin{equation}}
\newcommand{\eeq}{\end{equation}}
\newcommand{\bea}{\begin{eqnarray}}
\newcommand{\eea}{\end{eqnarray}}
\def\plumin{\underline{+}}
\def\minplu{{{\stackrel{\underline{\ \ }}{+}}}}
\def\bfr{{\bf r}}
\def\bfw{{\bf w}}
\def\bfc{{\bf c}}

\def\hpsi{\hat \psi (\bfr)}
\def\hpsid{\hat \psi^\dagger (\bfr)}
\def\tpsi{\tilde \psi (\bfr)}
\def\tpsid{\tilde \psi^\dagger (\bfr)}

\def\hphi{\hat \phi (\bfr)}
\def\hphid{\hat \phi^\dagger (\bfr)}
\def\tphi{\tilde \phi (\bfr)}
\def\tphid{\tilde \phi^\dagger (\bfr)}

\def\Phir{\Phi (\bfr)}
\def\Phirs{\Phi^* (\bfr)}
\def\ntl{\tilde n}
\def\mtl{\tilde m}
\def\ntlr{\tilde n (\bfr)}
\def\mtlr{\tilde m (\bfr)}

\title{Novel Dynamical Resonances in Finite-Temperature Bose-Einstein Condensates}
\author[A. J. Geddes, S. A. Morgan and D. A. W. Hutchinson]
{A. J. Geddes$^1$, S. A. Morgan$^2$ and D. A. W. Hutchinson$^1$}

\affiliation{$^1$Department of Physics, University of Otago,
P.O. Box 56, Dunedin, New Zealand
\\$^2$Department of Physics and Astronomy, University College London, Gower Street,
London WC1E 6BT, United Kingdom}

\label{firstpage}



\maketitle

\begin{abstract} {Bose-Einstein condensation, collective excitations, thermal
field theory}

We describe a variety of intriguing mode-coupling effects which can occur in a
confined Bose-Einstein condensed system at finite temperature. These arise from
strong interactions between a condensate fluctuation and resonances of the
thermal cloud yielding strongly non-linear behaviour. We show how these
processes can be affected by altering the aspect ratio of the trap, thereby
changing the relevant mode-matching conditions. We illustrate how direct driving of the thermal cloud can lead
to significant shifts in the excitation spectrum for a number of modes and provide further experimental scenarios in which the dramatic behaviour observed for the $m=0$ mode at JILA (Jin {\it et al.} 1997) can be repeated. Our theoretical description is based on a successful second-order
finite-temperature quantum field theory which includes
the full coupled dynamics of the condensate and thermal cloud and all relevant
finite-size effects.
\end{abstract}


\section{Introduction}

Bose-Einstein condensation (BEC) in dilute atomic gases offers a virtually unique
laboratory for the investigation and experimental testing of thermal quantum field
theories (QFT). The ability to perform both {\it a priori} calculations and
precision experiments for direct comparison has led to a deepening of our
understanding of the important processes which must be included in a successful
finite-temperature QFT. The pioneering measurements of condensate excitation
frequencies and decay rates at the Joint Institute for Laboratory Astrophysics (JILA)
have provided the most stringent tests of such
theories to-date (Jin {\it et al.} 1997) and efforts to explain these results have led to the development of a second-order QFT which simultaneously includes both the dynamic coupling of the
condensed and uncondensed atoms and the effects of strong interactions and
finite-size (Morgan {\it et al.} 2003, Morgan 2004). Of particular importance is the fact that the theory explicitly includes the effect of direct driving of the thermal cloud by the perturbation which generates the excitations. This allows the condensate to be excited indirectly via the non-condensed atoms as well as directly via the perturbation, a process which can be very important at finite temperature when there is a large non-condensate fraction. The final agreement between the theoretical predictions and the JILA measurements is
excellent, with no adjustable parameters required in the theory (Morgan {\it et al.} 2003). This was the
first QFT calculation for confined BECs which systematically included the phonon
character of the low lying states, all relevant Landau and Beliaev processes and
the effects of finite system size.

In this paper we propose new spectroscopic experiments to further test this thermal QFT and to explore the intriguing mode-coupling effects which can occur in a trapped condensate at finite temperature. Initially
we consider a
condensate confined in a spherically symmetric harmonic trap. This geometry has been previously considered (Rusch {\it et al.} 2000, but in the absence of direct driving of the thermal cloud and by the Pisa group Liu {\it et al.} 2004) using a different formalism based upon the random phase approximation. The importance of this geometry is in the mode matching characteristics of the excitation spectrum. In a spherical trap, a condensate has a breathing mode excitation with zero angular momentum ($l=0$) and a frequency which lies just above a strong resonance of the thermal cloud. As a consequence, there is a strong coupling between the two with the result that the mode energy is almost impervious to temperature, increasing slightly as the system heats up in contrast with the substantial downward shifts predicted for all other modes studied to date and observed in experiment (Jin {\it et al.} 1997, Morgan {\it et al.} 2003, Hutchinson {\it et al.} 1997, Dodd {\it et al.} 1998, Giorgini 1998, Morago {\it et al.} 2001). In addition, at temperatures approaching the critical temperature, where the thermal cloud dynamics can become dominant, indirect excitation of the condensate via the thermal cloud as intermediary can lead to a dramatic shift in the $l=0$ excitation frequency. This is in direct analogy to the similar behaviour of the lowest-energy $m=0$ mode observed in the JILA TOP trap experiment which provoked much theoretical discussion before its final resolution (Morgan {\it et al.} 2003, Jackson \& Zaremba 2002). 

Since this strong coupling is a consequence of mode matching between condensate and thermal modes it should be possible to adjust this behaviour by tuning the condensate excitation frequency closer to that of the thermal cloud. This can be achieved in practice by altering the anisotropy of the trapping potential (Hodby {\it et al.} 2001). In particular, guided by the zero-temperature mode spectrum calculated by Hutchinson and Zaremba (1998), we identify a resonance between a low-lying $m=0$ condensate mode and the thermal cloud with a frequency $\omega= 2 \omega_r$ for a trap with an aspect ratio of $\omega_z/\omega_r=0.764$, where $\omega_r$ and $\omega_z$ are radial and axial trap frequencies respectively. We apply our formalism in this geometry and demonstrate the interesting phenomena which can be expected in this case.

Observation of the predicted effects in a spherical trap, and particularly in this resonant case, would provide additional confirmation of the validity of our theoretical description and intriguing further examples of the coupled dynamics of condensed and uncondensed atoms which can occur in trapped Bose gases at finite temperature.

\section{Theory}

Briefly,(Morgan {\it et al.} 2003, Morgan 2004) we begin from the generalized Gross-Pitaevskii equation (GPE) for a condensate of $N_0$ atoms 
\bea
i \hbar {\partial \Phi \over \partial t} & = & [\hat H _{sp} + P(\bfr,t) - \lambda (t) + N_0 (t) U_0|\Phi|^2 ]\Phi {}
\nonumber\\
& & {} + 2U_0 \ntl (\bfr,t)\Phi + U_0 \mtl (\bfr,t) \Phi^* -f(\bfr,t),
\label{TDGGPE}
\eea
where $\hat H_{sp} = - {\hbar ^2 \nabla^2 \over 2m} + V_{ext}(\bfr)$ is the single particle Hamiltonian, containing the kinetic energy and the static trap potential, $V_{ext}(\bfr)$. We take $V_{ext}(\bfr)$ to be the usual cylindrically symmetric harmonic potential $V_{ext}(\bfr) = \frac{1}{2}m(\omega_r^2r^2+\omega_z^2z^2)$ with $r^2 = x^2+y^2$. The other terms in Eq.~(\ref{TDGGPE}) are as follows: $P(\bfr,t)$ is a time-dependent external perturbation which generates excitations, $\lambda(t)$ is a scalar which controls the condensate phase, $\tilde n(\bfr,t)$ is the non-condensate density, $\tilde m(\bfr,t)$ is the so-called anomalous average and $f(\bfr,t)$ is a 
term resulting from particle number conservation. Interactions are described by the usual contact potential with interaction strength $U_0 = 4\pi\hbar^2a_s/m$ where $a_s$ is the \textit{s}-wave scattering length. This means that our definition of $\tilde m(\bfr,t)$ is subject to the usual ultra-violet renormalization.

In the static case, the GPE has a time-independent solution which satisfies
\bea
[\hat H_{sp} - \lambda & + & N_0 U_0 |\Phir|^2 + 2U_0 \ntlr ]\Phir {}
\nonumber\\
& & {} + U_0 \mtlr \Phirs - f(\bfr) = 0,
\label{TIGPE}
\eea
where $\lambda$ is the condensate eigenvalue and is approximately equal to the chemical potential (Morgan 2000). The associated quasiparticle excitations obey a set of coupled Bogoliubov-de Gennes equations which we solve for the static quasiparticle wavefunctions $\{u_i(\bfr),v_i(\bfr)\}$ and energies $\{\epsilon_i\}$. We treat the thermal cloud as a non-interacting gas of quasiparticles coupled to the condensate mean-field with the result that all thermal cloud quantities can be calculated directly in terms of these quasiparticle solutions.

We now consider the effects of the external perturbation $P(\bfr,t)$, allowing all quantities to develop small time-dependent oscillations about their static values; 
$\Phi(\bfr,t)=\Phir + \delta \Phi(\bfr,t)$, $\ntl(\bfr,t) = \ntlr + \delta \ntl(\bfr,t)$ etc. Substituting into the time-dependent generalized GPE, Eq 
(\ref{TDGGPE}), and linearising we obtain the equation of motion for the condensate fluctuation $\delta \Phi(\bfr,t)$. This is solved by combining 
it with its complex conjugate, Fourier transforming and expanding in the static quasiparticle basis
\beq
 {\delta \Phi(\bfr,\omega) \choose \delta \Phi^*(\bfr,-\omega) } = \sum_i b_i(\omega) { u_i(\bfr) \choose v_i(\bfr) }.
\label{qb}
\eeq
The $b_i(\omega)$ expansion coefficients are directly related to the experimentally measured condensate density fluctuations, given by $\delta n_c = \delta(N_0 |\Phi|^2)$. 

Assuming that only a single condensate mode ``p'' is excited, the expansion coefficients are of the form 
\beq
b_p(\omega) = P_{p0}(\omega)G_p(\omega + i\gamma),
\eeq
where $P_{p0}(\omega)$ is the matrix element for the generation of the excitation from the condensate and $i\gamma$ is a small imaginary part in the frequency (see later). The resolvent $G_p(\omega)$ is defined as
\beq
G_p(\omega) = {1 \over \hbar\omega - \epsilon_p-\Sigma_p(\omega)},
\label{G_p}
\eeq
where $\Sigma_p(\omega)$ is a frequency-dependent self-energy given by
\beq
\Sigma_p(\omega) = \Delta E_{pp}^{(S)}+\Delta E_{pp}^{(D)}(\omega).
\eeq
$\Sigma_p(\omega)$ contains both static (S) and dynamic (D) contributions to the energy. The static part, $\Delta E_{pp}^{(S)}$, describes interactions between the condensate and the equilibrium non-condensate 
mean fields, while the frequency-dependent dynamic part $\Delta E_{pp}^{(D)}(\omega)$ describes the driving of the thermal cloud by the condensate fluctuation and the subsequent back-action.

So far we have assumed that the external perturbation only directly excites the condensate. All the dynamics of the thermal gas is therefore due to the subsequent action of the condensate on the uncondensed cloud. A second possibility exists, however, in which the external perturbation also directly drives the non-condensate. Including this process, we find that the expansion coefficients are given by
\beq
b_p(\omega) = P_{p0}(\omega)R_p(\omega + i\gamma),
\eeq
where the new resolvent $R_p(\omega)$ is defined by 
\beq
R_p(\omega) =\left[ 1+{\Delta P_{p0}^{(S)} + \Delta P_{p0}^{(D)}(\omega) \over P_{p0}(\omega)} \right]G_p(\omega).
\label{R_p}
\eeq
Again there are both static and dynamic contributions present in this expression. The static term $\Delta P_{p0}^{(S)}$ is due to the effect of changes in the static condensate shape on the excitation matrix element, 
while the more interesting dynamic term is due to direct excitation of thermal cloud fluctuations by the external potential and their subsequent coupling to the condensate.

Both the dynamic terms $\Delta E_{pp}^{(D)}(\omega)$ and $\Delta P_{p0}^{(D)}(\omega)$ are calculated via a sum over all the Landau and Beliaev processes which can occur in the thermal cloud. These are resonant whenever certain energy matching criteria are satisfied and the parameter $\gamma$ is therefore required to keep the expressions finite at the corresponding poles. Physically, $\gamma$ describes the finite resolution of the experiment and typical values are of order a few times $10^{-2}$ of the trapping frequency. In this paper we have taken $\gamma = 0.05\omega_r$ but our results are not sensitive to its precise value within the experimentally relevant range.

\section{Spherical Geometry}\label{sect3}

\begin{figure}[h]
{\centering \resizebox*{!}{0.45\textheight}{\includegraphics{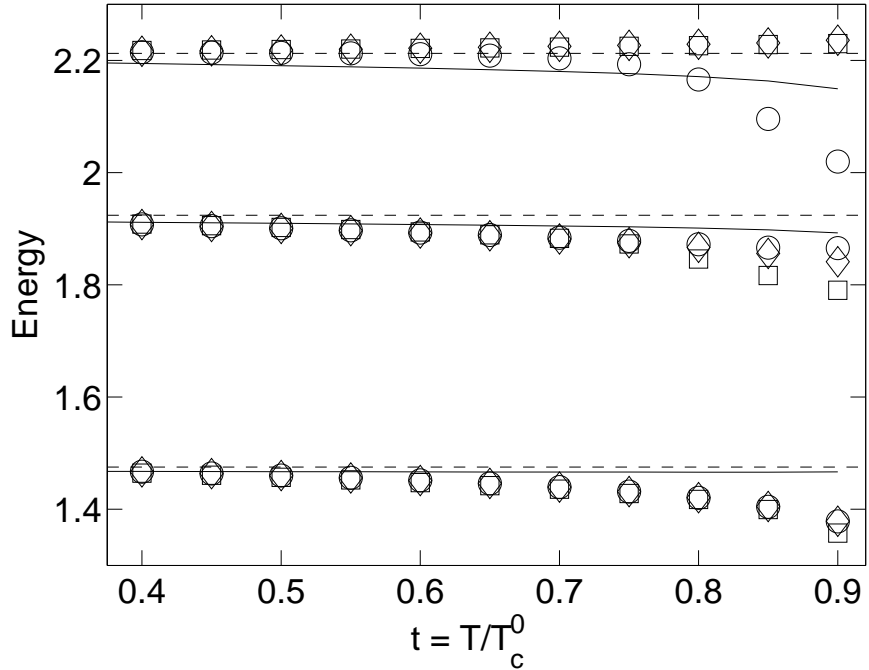}} \par}
\caption{\label{fig1} Excitation energies as a function of reduced temperature $t = T/T_c^0$ for the lowest-energy ($n = 1$, $l = 0$ ), ($n = 0$, $l = 3$) and ($n = 0$, $l = 2$) quasiparticle modes, from highest energy down. The diamonds and 
circles exclude (G) and include (R) thermal driving, respectively. Squares give the mean-value of the self-energy at the poles of the resolvent, the dashed lines are the unperturbed Bogoliubov energies and the solid lines are the results of a perturbative implementation of the HFB-Popov theory. Energies are in units of the radial trapping frequency.}
\end{figure}

Figure~\ref{fig1} shows results for the energies of three quasiparticle excitations in a spherical trap as a function of reduced temperature. The condensate contains $6000$ atoms and the trap frequency is $\omega_r = 2\pi\times 100$Hz. Results are shown from our second order QFT including and excluding the effect of direct thermal driving, as well as from a perturbative HFB-Popov calculation in which the thermal cloud is held static (Griffin 1996, Hutchinson {\it et al.} 1997, Hutchinson {\it et al.} 2000). The external perturbation was assumed to be a low-order polynomial multiplied by the spherical harmonic with the appropriate symmetry for each mode. For the $l=0$ and $l=2$ modes it therefore had a quadratic dependence on the co-ordinates, while for the $l=3$ mode it was a cubic function.

A number of features of Fig.~\ref{fig1} are of interest and demonstrate the importance of the dynamic coupling between the condensate and non-condensate. In general, the second order theory predicts markedly greater shifts than the HFB-Popov theory. For the $l=3$ and $l=2$ modes, this shift is negative at high temperature and direct driving of the non-condensate has little effect. In contrast, the $l=0$ mode is almost unaffected by temperature, shifting slightly upwards as the temperature rises, in contrast to the HFB-Popov prediction which shows a substantial downwards shift. For this mode, there is also a dramatic difference between the results obtained with and without direct driving of the thermal cloud. If this process is included, the response shifts towards the non-interacting gas value of $2\omega_r$ for temperatures greater than about $0.7T_c^0$, where $T_c^0$ is the ideal gas critical temperature. This is reminiscent of the similar shift seen for the $m=0$ mode in the TOP trap of the JILA experiment and the physical explanation is the same in this case; the condensate excitation is usually excited directly by the external perturbation but at high temperatures it can also be excited indirectly via the thermal cloud. If the thermal cloud response is particularly large at the frequencies of interest and if the coupling to the condensate is sufficiently strong, this second excitation method can become dominant and the condensate response becomes peaked around that of the non-condensate.

We can gain some insight into the effects of the coupling between the condensed and uncondensed atoms by considering a simple model in which the thermal cloud response is characterized by a single resonance at a frequency $\omega_0$. If the condensate has a Bogoliubov energy of $E_p = \hbar\omega_p$ and the (temperature-dependent) coupling strength to the thermal cloud is $g^2$ then the resolvent $\mathcal{G}_p(\omega)$ has the form
\beq
\hbar\mathcal{G}_p(\omega) = \frac{1}{\omega+i\gamma-\omega_p-\frac{g^2}{\omega+i\gamma-\omega_0}}.
\eeq
In the limit $g = 0$, this is the usual lorentzian with a single peak at the Bogoliubov frequency, $\omega = \omega_p$. For finite $g$ and in the limit $\gamma \rightarrow 0$, $\mathcal{G}_p$ has two poles at $\omega = (\omega_p + \omega_0)/2 \pm \sqrt{(\omega_p -\omega_0)^2/4 + g^2}$. Thus as $g$ increases the poles move further apart and in particular the original pole at $\omega \sim \omega_p$ is shifted away from the resonance at $\omega_0$. This is the usual level repulsion familiar from  second order perturbation theory.

\begin{figure}[h]
{\centering \resizebox*{!}{0.6\textheight}{\includegraphics{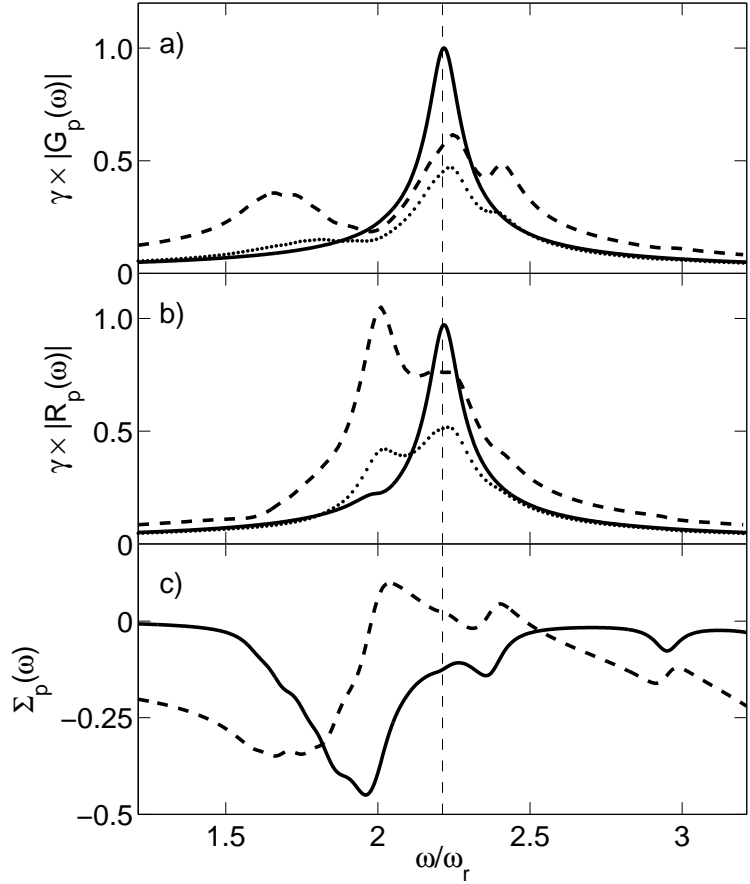}} \par}
\caption{\label{fig2} Response functions and self-energies for the $(n=1,l=0)$ mode in a spherical trap.
a) $\mathcal{G}_p(\omega)$ at t = 0, 0.75, 0.9 (solid, dotted, dashed (x2) respectively).
b) $\mathcal{R}_p(\omega)$ at t = 0, 0.75, 0.9 (solid, dotted, dashed (x2) respectively).
c) Real (dashed) and imaginary (solid) parts of the self-energy $\Sigma_p(\omega
)$ at t = 0.9.}
\end{figure}

If we include the effect of direct driving of the non-condensate with coupling strength $g_P$, the resolvent $\mathcal{G}_p(\omega)$ is replaced by $\mathcal{R}_p(\omega)$, which has the form
\beq
\hbar\mathcal{R}_p(\omega) = \frac{1+\frac{g_P g}{\omega+i\gamma-\omega_0}}{\omega+i\gamma-\omega_p-\frac{g^2}{\omega+i\gamma-\omega_0}}.
\eeq
If $g_P$ is small and/or the frequency difference $\omega_p-\omega_0$ is large, this does not differ greatly from $\mathcal{G}_p(\omega)$ for frequencies around $\omega_p$. However, if $g_P$ is large and $\omega_p-\omega_0$ is small, the additional factor in the numerator can dominate the response which becomes again a simple lorentzian but now centred on the thermal cloud resonance $\omega_0$. This cross-over behaviour can clearly be seen in Figs.~\ref{fig2}a and \ref{fig2}b where we plot the response functions $\mathcal{G}_p(\omega)$ and $\mathcal{R}_p(\omega)$ as a function of frequency for various reduced temperatures. The strong response at $\omega=2 \omega_r$ is visible in $\mathcal{R}_p(\omega)$ at $0.9 T_c^0$ but is absent in $\mathcal{G}_p(\omega)$ which actually has a local minimum in this region. At this temperature and frequency, the excitation of the condensate via the thermal cloud is more than four times as large as the direct effect of the perturbation.

The frequency-dependence of the real and imaginary parts of the self-energy $\Sigma_p(\omega)$ for this mode are shown in Fig.~\ref{fig2}c) at a temperature $T = 0.9 T_c^0$. Although this clearly has other structure, its form is rather close to that assumed in our analysis above, consisting of a (negative) constant and a single large resonance at a frequency $\omega = 2\omega_r$. The value of the negative constant for this case is of order $-0.13\hbar\omega_r$ and without the resonance structure the mode would shift downwards by about this amount. Comparison with Fig.~\ref{fig1} shows that such a shift is of the same order as the downwards shifts observed for the $l=3$ and $l = 2$ modes at this temperature. The reason the $l=0$ mode is instead almost impervious to temperature in the absence of direct thermal driving is therefore the result of strong level repulsion from the thermal cloud resonance at $\omega = 2\omega_r$. The upward shift this produces almost exactly counterbalances the downward shift we would otherwise obtain from the cumulative effect of all other modes in the problem. That this interpretation is correct can also be seen from the appearance of a second peak in the response function $\mathcal{G}_p(\omega)$ at $\omega \sim 1.7$ at $t = 0.9$ shown in Fig.~\ref{fig2}a. Comparing with our simple model above, we see that this comes from the second pole which originates at the thermal cloud resonance $\omega = 2\omega_r$ in the absence of interactions but is pushed down by level repulsion with the primary condensate resonance at $\omega \sim 2.2 \hbar \omega_r$.

The thermal cloud resonance at $\omega = 2\omega_r$ has its origin in a large number of Landau processes which correspond to removing an atom from a single-particle level with quantum numbers $(n_1,l_1)$ and placing it in a level with quantum numbers $(n_2,l_2)$. Since the energies of an ideal gas in a spherical trap are given by $\epsilon_{nl} = \hbar\omega_r(2n+l+3/2)$, these processes are resonant at frequencies $\omega_{12} = \omega_r(2(n_2-n_1)+l_2-l_1)$, so at high temperature the non-condensate has a large response at integer multiples of the trap frequency. In particular, there is a large response around $\omega = 2\omega_r$ corresponding to processes where $l_2-l_1$ is even. Since the perturbation required to excite a Bogoliubov mode with angular momentum $l_p$ involves the spherical harmonic $Y_{l_p}^{m_p}(\theta,\phi)$, we have the parity selection rule $l_p+l_1+l_2 = \mbox{even}$. Thus, if the condensate has a quasiparticle excitation with even angular momentum close in energy to an even integer multiple of the trap frequency we can expect its dynamics to be strongly influenced by the thermal cloud. This situation indeed occurs in a spherical trap where the lowest $l=0$ mode has an energy of order $2.2\hbar\omega_r$. The $l=3$ mode shows no such influence of the thermal cloud, despite having an energy significantly closer to $2\hbar\omega_r$, because it has odd parity and is therefore decoupled from the thermal cloud resonances in this region.

In a TOP trap a similar situation arises because there is a mode with even z-parity and zero angular momentum about the z-axis ($m = 0$) with a Bogoliubov energy of order $1.8\hbar\omega_r$. This mode is therefore strongly coupled to Landau resonances with frequencies around $2\omega_r$ and the same effect of direct driving of the non-condensate is observed. In this case, however, the condensate mode lies below the thermal cloud resonance with the result that the effect of thermal cloud driving is to shift its frequency up towards the non-interacting value, in contrast to the downwards shift this process produces for the $l=0$ mode in a spherical trap. This difference also shows up in the relative phase of the condensate--non-condensate oscillations. When the condensate drives the thermal cloud these oscillations are generally out-of-phase at high temperature, but in the reverse situation the relative phase depends on whether the thermal cloud resonance is above or below the unperturbed condensate mode. For the $m=0$ mode at JILA this resonance is above the primary condensate energy and the growing importance of direct thermal driving is associated with a transition to in-phase oscillations of the two clouds. In contrast, in the spherical trap the thermal cloud resonance is below the original condensate mode and we find that the oscillations remain predominately out-of-phase at the highest temperatures.

We note that whereas our results for the self-energy and resolvent $\mathcal{G}_p(\omega)$ are intrinsic to the system, depending only on the geometry and particle interactions, the effects of direct thermal driving are sensitive to the particular form of the perturbation used to excite the system and so will vary from one experiment to another. This provides an important handle on the effects we have discussed since it should be possible to vary the strength of direct thermal driving by modifying the form of the perturbation. For example, it should be possible to design an experiment to probe our predictions based upon either a localised perturbation of the magnetic trap or a Raman scheme. The localised perturbation could be designed to couple predominantly to the condensate, which occupies the centre of the trap, rather than to the non-condensate which is peaked nearer the edges of the atomic cloud. A Raman based scheme could be designed to be resonant with the condensate and not the non-condensate, thus directly exciting only the condensate atoms. Such experiments would allow the effects of direct thermal driving to be separated from other self-energy interactions and give access to the intrinsic response function $\mathcal{G}_p(\omega)$.

\section{Resonant Geometry}

\begin{figure}[h]
{\centering \resizebox*{!}{0.45\textheight}{\includegraphics{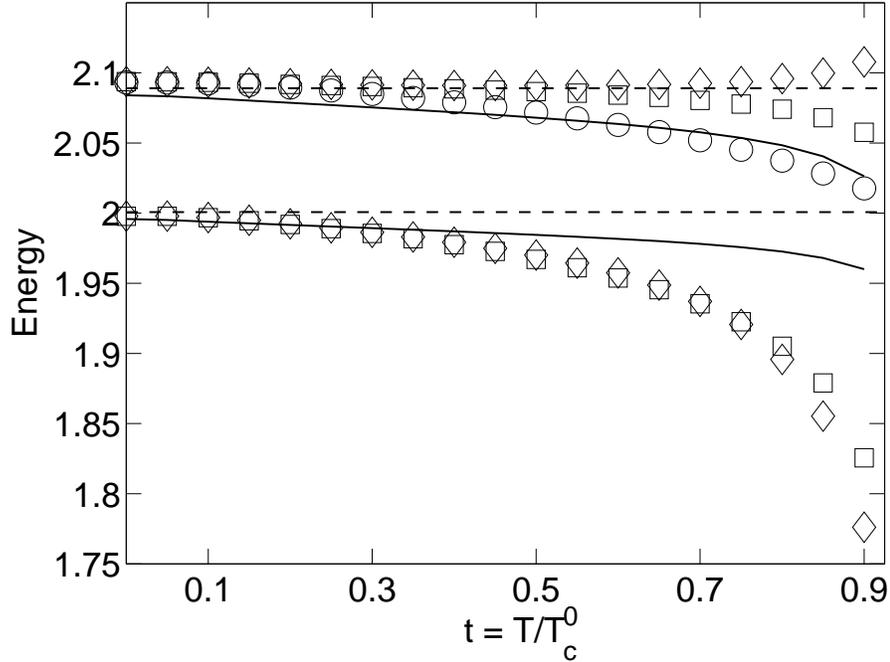}} \par}
\caption{\label{fig3} Excitation energies for the $(m,p,n) = (0,0,2)$ (lower) and $(0,0,3)$ (upper) modes in an anisotropic trap as a function of reduced temperature. Symbols are as in Fig. 1.}
\end{figure}

To explore mode matching effects further, we can tune the condensate mode into resonance with a thermal cloud excitation frequency by adjusting the aspect ratio of the trap. This is experimentally feasible and has been used to explore nonlinear effects such as parametric down-conversion in the past (Hodby {\it et al.} 2001). By selecting an aspect ratio of $\omega_z/\omega_r=0.764$ we are able to tune the zero temperature, Bogoliubov frequency for one of the low lying $m=0$ modes to be precisely $2 \omega_r$ (this mode corresponds to the quantum numbers $(m,p,n) = (0,0,2)$ where $m$ is the angular momentum about the \textit{z}-axis, $p$ is the \textit{z}-parity ($0$ is even) and $n$ indicates the energy in this subspace). With this geometry we repeat the calculation and again see shifts in the excitation frequencies of low-lying modes with increasing temperature as displayed in Fig.~\ref{fig3}.

It is interesting to note that when we tune the $(0,0,2)$ mode into resonance, due to a fortuitous anti-crossing behaviour, the $(0,0,3)$ mode is relatively close in frequency. We therefore see an additional strong coupling effect for this mode. In fact we find that because of the change in matrix elements associated with the change in trap geometry, this mode couples more strongly to the thermal cloud resonance at $\omega = 2\omega_r$ than the mode tuned to resonance. Its behaviour is therefore very similar to that previously observed for the $l=0$ mode in the spherical trap and plots of the response functions $\mathcal{G}_p(\omega)$ and $\mathcal{R}_p(\omega)$ and self-energy $\Sigma_p(\omega)$ for this mode are qualitatively similar to those given in Fig.\ref{fig2}.

\begin{figure}[h]
{\centering \resizebox*{!}{0.40\textheight}{\includegraphics{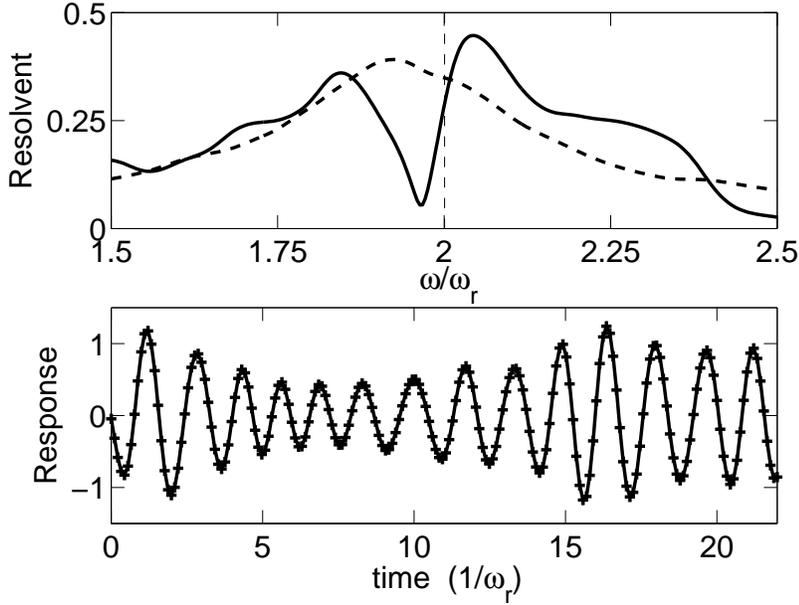}} \par}
\caption{\label{fig4} Upper: $\gamma \times |\mathcal{G}_p(\omega)|$ (dashed) and $\gamma \times |\mathcal{R}_p(\omega)|$ (solid) vs frequency at reduced temperature $t = 0.7$ for the (0,0,2) mode in a resonant trap. The vertical dashed line shows the position of the original unperturbed Bogoliubov mode.
Lower: Plot of the condensate density response (arbitrary units) in the time domain at $t = 0.7$ including the effect of direct driving of the non-condensate (corresponds to the solid line above). Collapse and revival of the condensate oscillations are clearly visible as a result of the split of the spectrum into two distinct parts.}
\end{figure}

In contrast, the coupling in the self-energy for the $(0,0,2)$ mode is weaker (although still quite strong) with the result that this mode has the usual strong downwards shift in energy in the absence of direct thermal driving. The mode is nonetheless strongly affected by direct excitation of the thermal cloud, however, so when this process is included we get a strong component to the response around $\omega = 2\omega_r$. In contrast with the other cases we have examined, this now causes the response to split into two distinct peaks. This is shown in the upper panel of Fig.~\ref{fig4} where the resolvents, $\mathcal{G}_p(\omega)$ and $\mathcal{R}_p(\omega)$ are plotted at a temperature of $0.7 T_c^0$. The two peak structure comes from the product of the underlying lorentzian function for $\mathcal{G}_p(\omega)$ centred near $\omega=1.9 \omega_r$ and the thermal cloud resonance at $2.0 \omega_r$ which modifies $\mathcal{G}_p(\omega)$ to $\mathcal{R}_p(\omega)$. The product of these two misaligned peaks yields the bimodal structure seen in Fig.~\ref{fig4}. Note that since this bimodal distribution renders a lorentzian fit meaningless, a single frequency cannot be ascribed so this is omitted in Fig.~\ref{fig3}. 

A double-peak response of this form is a signature of collapse and revival between coupled modes. This is most clearly seen in the time domain shown in the lower panel of Fig.~\ref{fig4}. Here we have plotted the condensate response for the $(0,0,2)$ mode as a function of time for the same temperature as in the upper panel. Initially the oscillations decay, but after approximately ten trap periods the amplitude increases again as the mode intensity is transfered back from the thermal cloud. We should note that the coupling here is to a resonance formed from the cooperative effect of many Landau processes. This means that damping could become a significant issue, although our estimates of the damping rate indicate that this collapse and revival should be observable on the time scales of typical experiments.

The fact that this geometry produces two modes with the same symmetry close in energy and both strongly coupled to a thermal cloud resonance means that in principle it may be possible to see interesting new effects in this case where energy is transfered between the two modes via the thermal cloud as mediator. Such processes are included in the QFT we have developed (Morgan 2004) but require a significant extension of our numerical methods and are therefore not contained in the present calculation. Such effects along with those described in this paper are part of the rich dynamical structure which exists in condensed systems at finite temperature and which is now accessible to detailed experimental and theoretical study. 

\section{Conclusions}

In conclusion, we have presented concrete predictions for the excitation energies of low-lying modes of a Bose-condensate in a spherical trap at finite temperature using a successful second order QFT which includes the dynamic coupling of condensed and uncondensed atoms and finite size effects. For the lowest breathing mode with zero angular momentum, we have observed the effects of strong coupling between the condensate and thermal cloud. This leads to a mode energy which is almost independent of temperature, shifting only slightly upwards, compared with the substantial downward shifts predicted and observed in previous calculations and experiments. In addition, at high temperature we observe a strong crossover to the non-interacting limit which is a consequence of indirect excitation of the condensate via the thermal cloud as intermediary, in direct analogy to the similar phenomenon observed in the JILA TOP trap. The observation of these effects would provide further confirmation of the validity of our theoretical approach and of the importance of the dynamic coupling between the condensate and thermal cloud in trapped gas BECs at finite temperature.

We have also identified a new regime where, by altering the aspect ratio of the trap, it is possible to bring condensate modes into resonance with modes of the thermal cloud. Strong mode coupling, especially when the thermal cloud is directly driven by the external perturbation, is then observed allowing the observation of non-linear phenomena such as collapse and revival between condensate and thermal cloud modes. We also note that, due to a fortuitous anti-crossing in the condensate mode spectrum, another mode with the same symmetry is also coupled to the thermal cloud resonance in the vicinity of $\omega=2 \omega_r$. This should allow for a coupling between condensate modes mediated via the non-condensate dynamics which would prove a very interesting subject for future theoretical and experimental investigation.

\begin{acknowledgements}
We thank the Royal Society (London), the Marsden Fund of the Royal Society of New Zealand and the University of Otago for financial support for this project.
\end{acknowledgements}




\end{document}